\documentstyle[aps,prl,preprint,floats,epsfig]{revtex}

\def\LCP{\Lambda_c^{+}}

\def\LA{\Lambda}

\def\EP{e^{+}}
\def\EM{e^{-}}

\def\To{\rightarrow}

\def\CASCZ{\Xi_c^{0}}

\def\CASCZHD{\Xi_c^{0} \To \Xi^{-} \pi^+}
\def\CASMIN{\Xi^- \To \LA \pi^-}

\def\EPEM{e^{+} e^{-}}
\def\SLOPEXA{-0.291 \pm 0.021\mbox{(stat)}^{+0.019}_{-0.015}\mbox{(syst)}}

\def\SLOPEAA{0.26 \pm 0.18\mbox{(stat)}^{+0.05}_{-0.04}\mbox{(syst)}}

\def\SLOPEBA{-0.57 \pm 0.21\mbox{(stat)}^{+0.05}_{-0.02}\mbox{(syst)}}

\def\AXCA{\alpha_{\CASCZ} = -0.56 \pm 0.39\mbox{(stat)}^{+0.10}_{-0.09}\mbox{(syst)}}

\begin{document}

\draft  

\preprint{\tighten\vbox{\hbox{\hfil CLNS 98/1586}
                        \hbox{\hfil CLEO 98-15}
}}

\title{A Measurement of the Decay Asymmetry Parameters in
{\boldmath $\CASCZHD$}}

\author{CLEO Collaboration}
\date{\today}

\maketitle
\tighten

\date{\today}
\maketitle

\begin{abstract} 
  Using the CLEO II detector at the Cornell Electron Storage Ring we
  have measured the $\CASCZ$ decay asymmetry parameter in the decay
  $\CASCZHD$. We find $\alpha_{\CASCZ} \alpha_{\Xi} = \SLOPEAA$, using
  the world average value of $\alpha_{\Xi} = -0.456 \pm 0.014$ we obtain
  $\AXCA$.  The physically allowed range of a decay asymmetry parameter
  is $-1<\alpha<+1$. Our result prefers a negative value:
  $\alpha_{\CASCZ}$ is $<0.1$ at the 90\% CL. The central value occupies
  the middle of the theoretically expected range but is not yet precise
  enough to choose between models.
\end{abstract}
\newpage

{
\renewcommand{\thefootnote}{\fnsymbol{footnote}}

\begin{center}
S.~Chan,$^{1}$ G.~Eigen,$^{1}$ E.~Lipeles,$^{1}$
J.~S.~Miller,$^{1}$ M.~Schmidtler,$^{1}$ A.~Shapiro,$^{1}$
W.~M.~Sun,$^{1}$ J.~Urheim,$^{1}$ A.~J.~Weinstein,$^{1}$
F.~W\"{u}rthwein,$^{1}$
% ...
\end{center}
 
\small
\begin{center}
$^{1}${California Institute of Technology, Pasadena, California 91125}\\
% ...
\end{center}

\setcounter{footnote}{0}
}
\newpage

The weak interactions underlying hadronic charm quark weak decay are
straightforward to describe theoretically, but complications arise
because the quarks are bound inside hadrons by the strong force.  These
interactions, which are described by the theory of quantum
chromodynamics (QCD), are very difficult to predict using perturbative
methods because the strong coupling is large at the typical energies of
charm decays.  Compared to charm mesons, charm baryons offer new
information for two reasons: first, non-factorizable contributions to
the decay amplitude are important; W-exchange diagrams can contribute
without the helicity suppression that decreases their contribution to
pseudoscalar meson decays, and internal W emission is expected to be
significant. The relative importance of these effects is believed to be
responsible for both the observed lifetime hierarchy of the charm
baryons and the relatively shorter lifetime of charm baryons compared to
charm mesons.  Second, parity violation in charm baryon decays is
readily observable because the decay of the daughter hyperon also
violates parity. In consequence a variety of models ~\cite{THEORY1} are
used to predict both the decay rate and degree of parity violation in
charm baryon decays. To constrain the models it is important to provide
as much experimental information as possible.

Parity violation occurs in hyperon and charm baryon ${1 \over 2}^+ \To
{1 \over 2}^+ 0$ decays due to the existence of two orbital angular
momentum decay amplitudes of opposite parity. The experimental
observable is an asymmetry in the angular decay distribution due to
interference between the two amplitudes.
In the decays $\LCP \To \LA \pi^+$ and $\CASMIN$ both followed by $\LA
\To p \pi^-$, the $\LA$ is produced with a polarization equal to 
\begin{equation}
 {\bf P_{\LA}} = { (\alpha_B + \hat{\bf \LA} \cdot {\bf P_B })\hat{\bf
\LA} - \beta_B ( \hat{\bf \LA} \times {\bf P_B} ) - \gamma_B \hat{\bf
\LA} \times ( \hat{\bf \LA} \times {\bf P_B} ) \over ( 1 + \alpha_B
\hat{\bf \LA} \cdot {\bf P_B} ) }
\label{polarize}
\end{equation}
where $\bf P_B$ is the parent baryon polarization, $\alpha_B ,~\beta_B$
and $~\gamma_B$ are the parent baryon asymmetry parameters and ${\bf
  \hat{\LA}}$ is a unit vector along the $\LA$ momentum in the parent
baryon frame ~\cite{KALLEN}. If the parent baryon polarization is
unobserved, or if the parent baryon is not polarized, Equation
~\ref{polarize} reduces to ${\bf P_{\LA}} = \alpha_B {\bf \hat\LA }$.
The angular distribution of a proton from the decay of a $\LA$ is
therefore
\begin{equation}
{dN \over d\cos \theta_{\LA}} \propto 1 + \alpha_{B} \alpha_{\LA} \cos
\theta_{\LA}
\label{lamasym}
\end{equation}
where $\theta_{\LA}$ is the angle between the proton momentum vector in
the $\LA$ rest frame and ${\bf \hat \LA}$, and $\alpha_{\LA}$ is the
$\LA$ decay asymmetry, which measures the degree of parity
violation in $\LA$ decay.  Experimentally, a fit to equation ~\ref{lamasym}
determines a product of decay asymmetries.  For example, in $\CASMIN$,
$\alpha_{B} = \alpha_{\Xi^-}$; since $\alpha_{\LA}$ is well measured
($\alpha_{\LA}=0.642 \pm 0.013$ ~\cite{PDG} ), $\alpha_{\Xi^-}$ can be
extracted.

Equation ~\ref{lamasym} has also been used to extract the decay
asymmetry of the $\Lambda_c$ in several decay modes.  For example in
$\LCP \To \LA \pi^+$, $\alpha_{B} = \alpha_{\LCP}$.  In the massless
fermion limit the chirality of the weak interaction corresponds to
$\alpha=-1$, however, the interplay between the strong and weak
interactions in the weak decay alters the value of $\alpha$.  The decay
asymmetry in $\LCP \To \LA \pi^+$ is relatively well
measured, the world average value is 
$\alpha_{\LCP}=-0.98 \pm 0.19$~\cite{PDG}, which is
consistent with the naive V-A expectation.  This result had been
predicted by Heavy Quark Effective Theory (HQET) in conjunction with the
factorization hypothesis~\cite{BJORKEN}. Most pole and quark
models~\cite{THEORY1}, also predict large negative values of
$\alpha_{\LCP}$. 

\begin{figure}[!h]
\centerline{ \psfig{figure=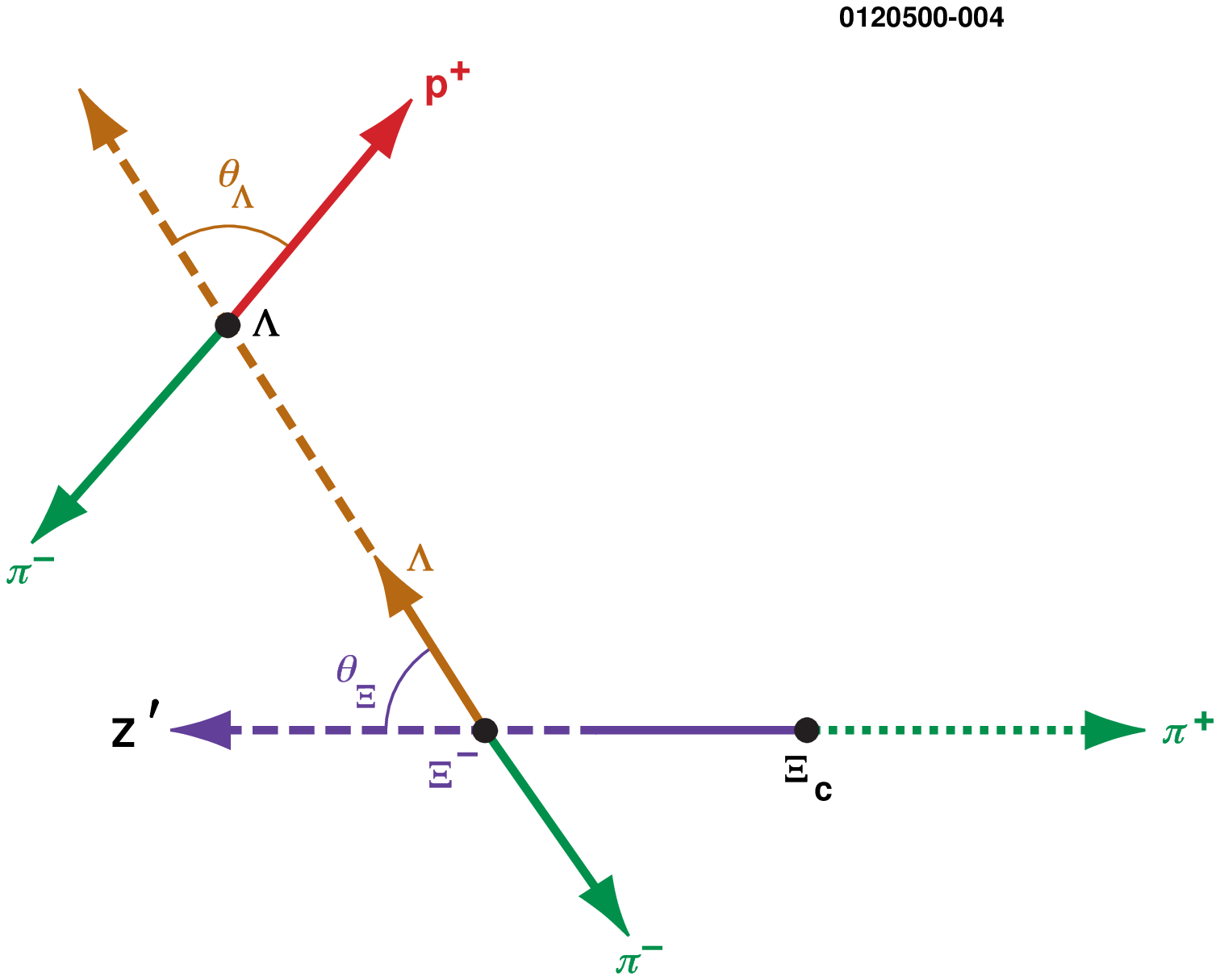,width=0.4\textwidth} }
\caption{Definition of the angles $\theta_\Xi$ and $\theta_\LA$ in
$\CASCZHD$. The decay of each particle is drawn in its own rest frame.}
\label{alphacas}
\end{figure}

In this letter we measure the $\CASCZ$ decay asymmetry
parameter in the decay $\CASCZHD$ for the first time.  For $\CASCZHD$
theoretical predictions cover a larger range: $-1 < \alpha_{\CASCZ}
<-0.38$ ~\cite{THEORY1}.
The decay of $\CASCZ$ is a three step process where parity violation
occurs at each decay stage: $\CASCZHD$, $\Xi^- \To \LA \pi^-$, $\LA \To
p \pi^-$.  The differential rate is therefore given by~\cite{PRIVAT}:
\begin{eqnarray}
\nonumber {dN \over d\cos \theta_{\Xi} d\cos \theta_{\LA}} \propto 1 & +
& \alpha_{\CASCZ} \alpha_{\Xi} \cos \theta_{\Xi} + \alpha_{\LA}
\alpha_{\Xi} \cos \theta_{\LA} \\ & + &\alpha_{\LA} \alpha_{\CASCZ} \cos
\theta_{\Xi} \cos \theta_{\LA}
\label{fullasym}
\end{eqnarray}
where in addition to the terms previously defined: $\theta_{\Xi}$ is the
angle between the $\LA$ momentum vector in the $\Xi^-$ rest frame and
the $\Xi^-$ momentum vector in the $\CASCZ$ rest frame as shown in
Fig.~\ref{alphacas}. Equation ~\ref{fullasym} reduces to the familiar
form of Equation ~\ref{lamasym} when integrated over $\cos
\theta_{\LA}$.

The data sample in this study was collected with the CLEO II detector
~\cite{KUBOTA} at the Cornell Electron Storage Ring (CESR).  The
integrated luminosity consists of 4.83 $fb^{-1}$ taken at and just below
the $\Upsilon (4S)$ resonance, corresponding to approximately 5 million
$e^+e^-\rightarrow c\overline{c}$ events.

We search for the decay $\CASCZHD$ in $\EPEM \To c \bar{c}$ events by
reconstructing the decay chain $\CASCZHD$, 
$\CASMIN$, $\LA \To p \pi^-$\footnote{Unless otherwise
  noted, throughout this letter charge conjugation is implied}.

The $\LA$ is reconstructed by requiring two
oppositely charged tracks to originate from a common vertex.
The positive track is
required to be consistent with a proton hypothesis\footnote{Hadronic particles are 
identified by requiring specific ionization energy loss measurements ($dE/dx$), combined 
with time-of-flight (TOF) information when available. The two measurements are combined 
into a joint probability for the particle 
to be a pion, a kaon or a proton. A charged track is defined to be consistent with a
particle hypothesis if its probability is greater than 0.003.}.
The momentum of the
$\LA$ candidate is calculated by extrapolating the charged track momenta
to the secondary vertex.  The invariant mass of $\LA$ candidates is
required to be within three standard deviations (3$\sigma$ = 6.0
MeV/c$^2$) of the known $\LA$ mass.  Track combinations which
satisfy interpretation as $K_s^0 \To \pi^+ \pi^-$ are rejected.
Combinatoric and $B$ decay backgrounds are reduced by requiring 
the momentum of $\LA$ candidates
be greater than 800 MeV/c. 

The $\Xi^-$ is reconstructed in the $\LA \pi^-$ decay mode.  $\Xi^{-}$
candidates are formed by combining each $\LA$ candidate with a
negatively charged track consistent with a pion hypothesis.  The $\Xi^-$
candidate vertex is formed from the intersection of the $\LA$ momentum
vector and the negatively charged track.  To obtain the $\Xi^-$
momentum, and $\LA \pi$ invariant mass, the momentum of the charged 
track is recalculated at the new vertex. The invariant $\LA \pi^-$ mass 
is fit to a double Gaussian signal shape
with parameters fixed from a GEANT ~\cite{GEANT} based Monte Carlo (MC)
simulation of the detector, and a first order Chebyshev polynomial to
describe the combinatorial background.  We find $8434 \pm 109$ events
consistent with $\Xi^- \rightarrow \LA \pi^-$ with an invariant mass of
$1322.0 \pm 0.03$ MeV/c$^2$ and width $\sigma_{av} = 2.5$ MeV/c$^2$,
where $\sigma_{av}$ is the weighted average width of the double
Gaussian. The mean and width are in agreement with the MC simulation.

\begin{figure}
\centerline{ \psfig{file=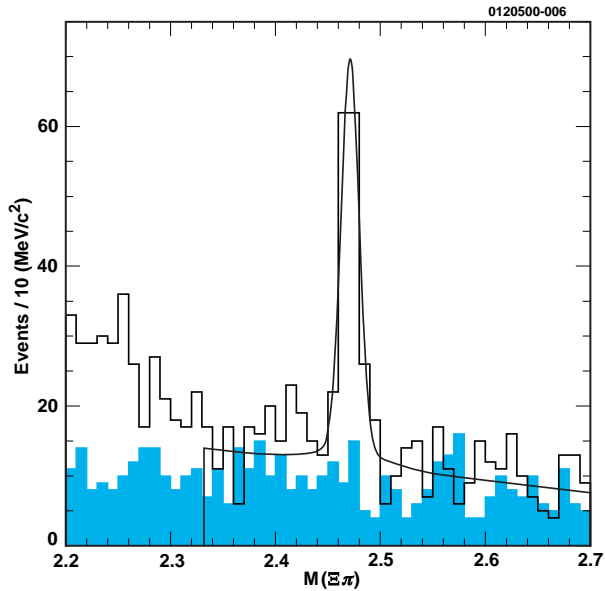,width=0.5\textwidth} }
\caption{Invariant $\Xi^- \pi^+$ mass distribution.  The shaded
histogram is wrong sign (WS) random $\Xi^- \pi^-$ combinations. }
\label{casc0}
\end{figure}

To reconstruct $\CASCZ$ candidates all $\Xi^-$ candidates are combined
with positively charged tracks consistent with a pion hypothesis. Since
charm fragmentation is a relatively hard process, the $\CASCZ$ spectrum
from $\EP \EM \To c \bar{c}$ is also fairly stiff. We therefore use a
scaled momentum cut, $x_p > 0.5$ \footnote{$x_p = p / p_{max}= p /
  \sqrt{E_{beam}^2 - m_{\CASCZ}^2}$ } to reduce combinatoric background.
This cut also eliminates charmed baryons from decays of B mesons.  Fig.
~\ref{casc0} shows the invariant mass distribution of $\Xi^- \pi^+$
combinations. The $\Xi^- \pi^+$ invariant mass distribution is fit to a
double Gaussian to describe the signal and a first order Chebyshev
polynomial to describe the combinatorial background, where the
parameters of the double Gaussian are fixed by MC simulation.  The
shaded histogram is wrong sign (WS) random $\Xi^- \pi^-$ combinations.
There are less WS than right sign (RS) random combinations due to charge
conservation.
The excess of right sign events over wrong sign events below M($\Xi^-
\pi^+$) $=2.33$ GeV/c$^2$ is due to feedthrough from $\Xi_c^{+,0} \To
\Xi^- n(\pi), \ n>1,$ where one or more pions are missing. The
kinematic limit for feedthrough is $M(\Xi_c^0)-M(\pi) = 2.3304$
GeV/c$^2$.  To simplify the fit to M($\Xi^- \pi^+$) the feedthrough region
is excluded.  We find $138 \pm 14$ $\CASCZHD$ candidates with a mass of
$2470.6 \pm 1.0$ MeV/c$^2$ and width $\sigma_{av}=10$ MeV/c$^2$.  The
mean and width are in agreement with MC simulation.  We require that the
invariant mass of $\CASCZ$ candidates be within 30 MeV/c$^2$ (3
$\sigma_{av}$) of the known $\CASCZ$ mass.

MC simulation
shows that the difference between generated and reconstructed values of
$\cos \theta_\Xi$ and $\cos \theta_{\LA}$ has both Gaussian and
symmetric non-Gaussian components. The resolution in $\cos
\theta_{\Xi,\LA}$, $\sigma( \cos\theta_\Xi,\LA)$, defined to be the
average of the rms variance of the Gaussian and of the non-Gaussian
component, weighted by the relative normalizations of the two components
\footnote{For $\cos \theta_\Xi$ and $\cos \theta_\LA$, the
  non-Gaussian component comprises 60\% and 32\% of the distribution,
  respectively} is: $\sigma( \cos \theta_\Xi) = 0.014$ and $\sigma( \cos
\theta_{\LA}) = 0.030$.

The decay asymmetry parameter in $\CASCZHD$ is measured in a two
dimensional unbinned maximum loglikelihood fit to the two-fold decay
angular distribution of Equation ~\ref{fullasym}, in a manner similar to
~\cite{SCHMIDT}.  This technique enables a multi-dimensional likelihood
fit to be performed to variables modified by experimental acceptance and
resolution. The probability function of the signal, $\Gamma_s,$ is determined 
by generating one high statistics MC sample of $\CASCZHD$,
$\CASMIN$, $\LA \To p \pi^-$ with a known value of
$\alpha_{\CASCZ}$ and the world average values of $\alpha_{\Xi}$ and
$\alpha_{\LA}$. The generated events are processed through the
detector simulation, off-line analysis programs, and selection
criteria.  Using the generated angles, accepted MC events are
weighted by the ratio of the decay distribution for a trial value of
$\alpha_{\CASCZ}$ to that of the generated distribution. 
By such weighting, a likelihood may be evaluated for each data event for
trial values of $\alpha_{\CASCZ}$, and a fit performed.  The
probability for each event is determined by sampling $\Gamma_s$ 
using a search area centered on each data point.  The size of the area is
chosen such that the systematic effect from a finite search area is
small and the required number of MC events is not prohibitively high.


Background is incorporated into the fitting technique by constructing
the log-likelihood function:
\begin{equation}
\ln {\mathcal{L}} = \sum_{i=1}^{N} ln ( P_S \Gamma_S + P_B \Gamma_B )
\label{lleqn}
\end{equation}
where $N$ is the number of events in the signal region and $P_S$ and
$P_B$ are the probabilities that events in this region are signal and
background respectively.
The probability distribution of background in the signal region,
$\Gamma_B$, is determined from 
($\Xi^- \pi^+$) mass sidebands above and below the signal region. 
The sidebands are: 2.5004 $<$ M($\Xi^- \pi^+$)$<$2.6904 GeV/c$^2$ 
and 2.3304 $<$ M($\Xi^- \pi^+$)$<$2.4404 GeV/c$^2$.

CP conservation requires $\alpha_{\CASCZ} =
-\alpha_{\bar{\CASCZ} }$. As the number of signal events is small
the analysis is insensitive to the presence of CP violation, 
therefore CP is assumed to be
conserved. Since Equation~\ref{fullasym} depends on
the products $\alpha_{\CASCZ} \alpha_{\Xi}$, $\alpha_{\CASCZ}
\alpha_{\LA}$ and $\alpha_{\Xi} \alpha_{\LA}$ which have the same sign
for particle and anti-particle states, particle and anti-particle
distributions are combined.

The validity of the analysis procedure is determined by constructing
artificial data sets consisting of MC signal events, generated with
known asymmetries, and background events taken from data. The generated
asymmetry was varied over the range: $-1 < \alpha_{\CASCZ} < 0$, in all
cases the fit returns unbiased values of $\alpha_{\CASCZ}$.

\begin{figure}[!h]
\centerline{
  \psfig{figure=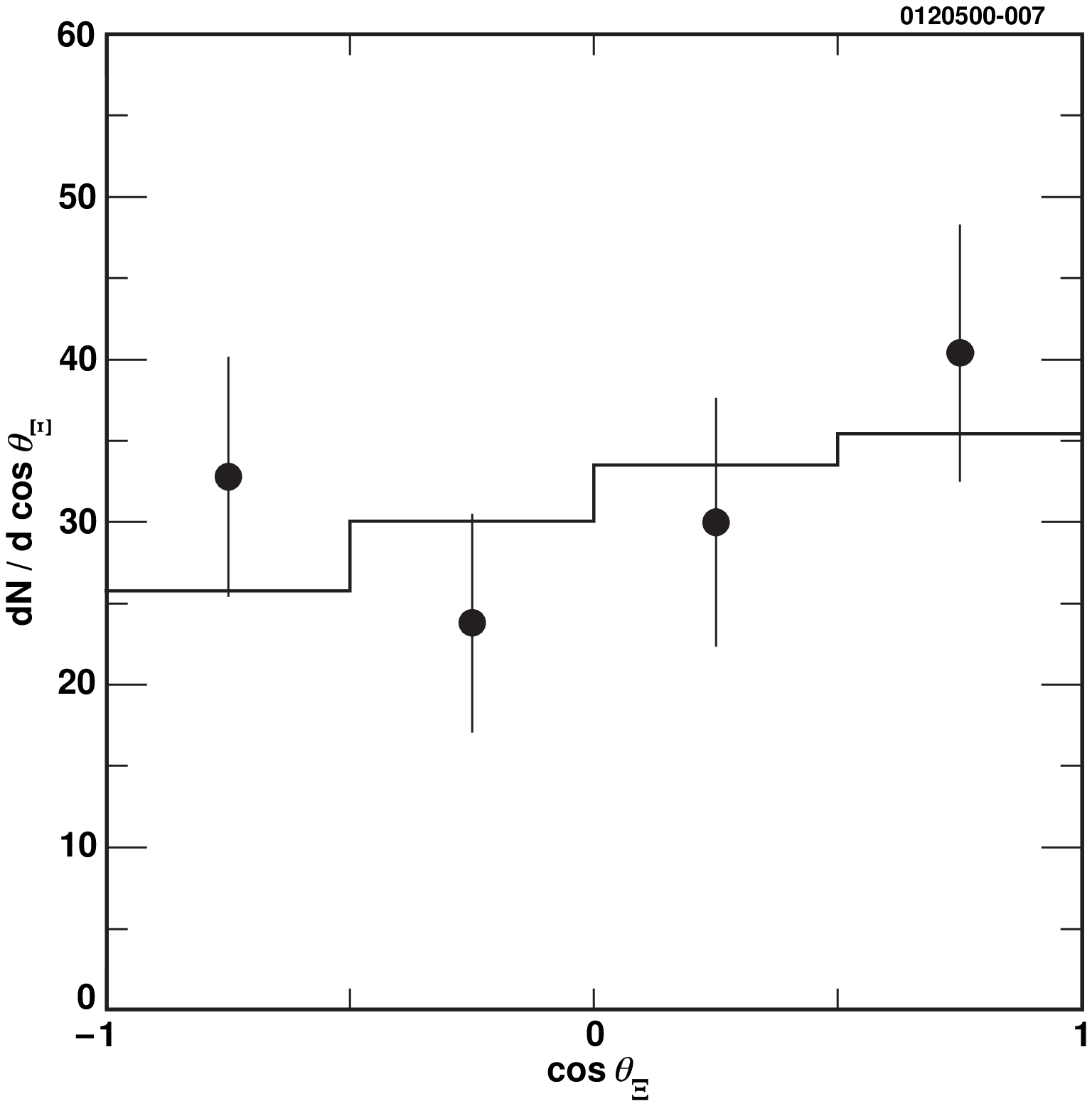,width=0.48\textwidth,height=2in}}
\centerline{
  \psfig{figure=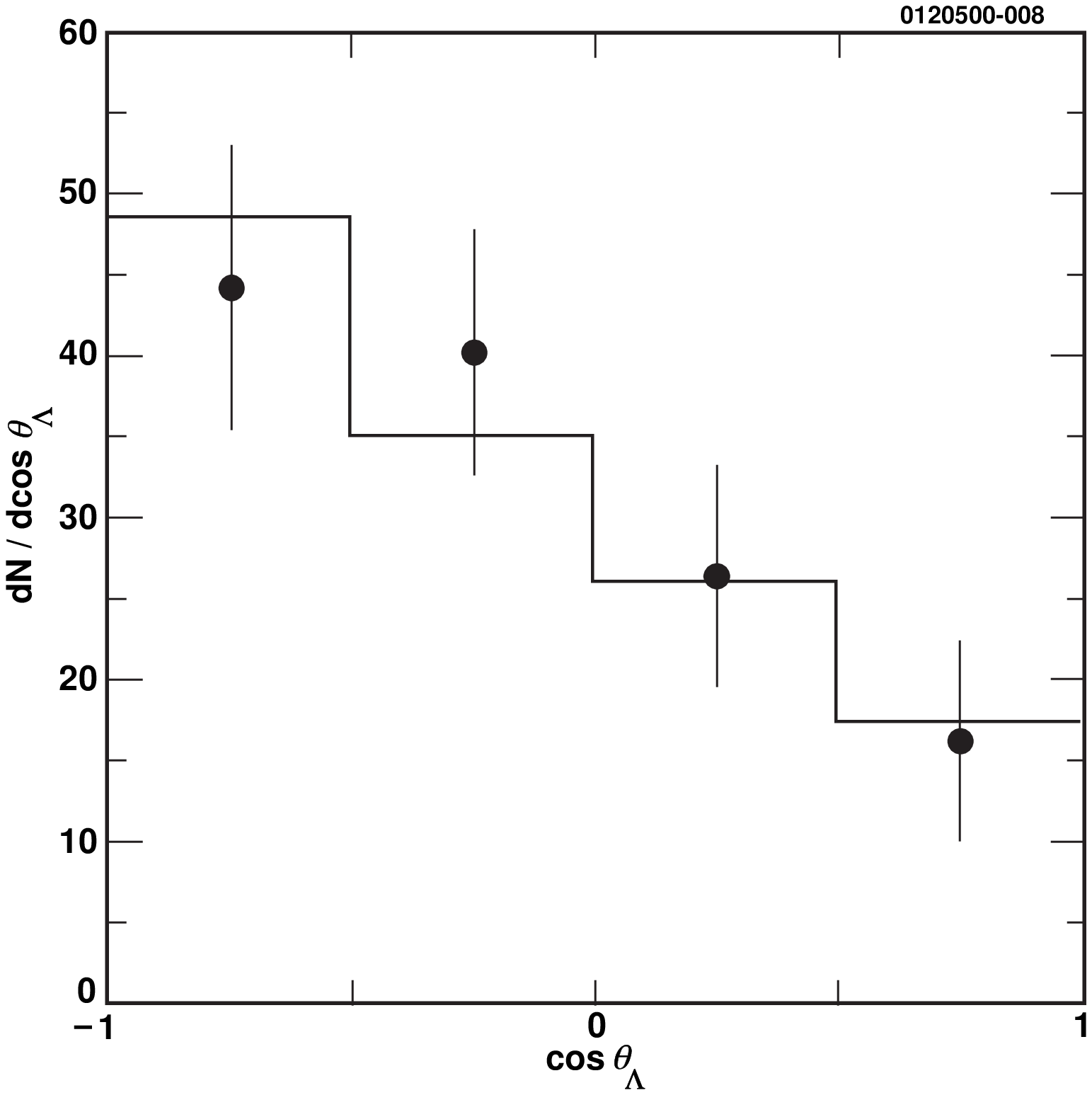,width=0.48\textwidth,height=2in}}
\caption{Upper plot: the cosine of the angle, $\theta_{\Xi}$, between
the $\LA$ momentum vector in the $\Xi^-$ rest frame
and the $\Xi^-$ momentum vector in the $\CASCZ$ rest frame, 
in the decay sequence $\CASCZHD$ $\CASMIN$, $\LA \To p \pi^-$ for 
sideband subtracted data 
(points with error
bars) and projection of the fit (solid histogram). Lower plot: the cosine of the angle,  
$\theta_{\LA}$, between the proton momentum vector in the $\LA$ rest frame and the 
$\LA$ momentum vector in the $\Xi^-$ rest frame in the decay sequence 
$\CASCZHD$ $\CASMIN$, $\LA \To p \pi^-$ for sideband subtracted data 
(points with error
bars) and projection of the fit (solid histogram).}  
\label{Xic6}
\end{figure}

Applying the fit procedure to the data we find $ \alpha_{\CASCZ}
\alpha_{\Xi} = 0.26 \pm 0.18 $, $\alpha_{\Xi} \alpha_{\LA} = -0.57 \pm
0.21 $ and $\alpha_{\CASCZ} \alpha_{\LA} = -0.26 \pm 0.37 $.  The
product $\alpha_{\CASCZ} \alpha_{\LA}$ is second order in decay angles,
consequently the fit is less sensitive to this quantity than to
$\alpha_{\CASCZ} \alpha_{\Xi}$ and $\alpha_{\Xi} \alpha_{\LA}$. We
provide a value of $\alpha_{\CASCZ} \alpha_{\LA}$ for completeness only.
The $\cos \theta_\Xi$ and $\cos
\theta_{\LA}$ distributions for data and projections of the fit
are shown in Fig.~\ref{Xic6}.
The value of $\alpha_{\Xi}\alpha_{\LA}$ found in the fit is in
reasonable agreement with a CLEO measurement of $\alpha_{\Xi}
\alpha_{\LA} = \SLOPEXA$ obtained with approximately 8,000 $\CASMIN$ events~\cite{ICHEP98}
and with the current world average value $\alpha_{\Xi} \alpha_{\LA} =
-0.293 \pm 0.007 $ ~\cite{PDG}. Constraining 
$\alpha_{\Xi}\alpha_{\LA}$ 
to the world average value does not reduce the statistical
error on $\alpha_{\CASCZ} \alpha_{\Xi}$. 

We have considered the following sources of systematic error and give
our estimate of the percentage error on $\alpha_{\CASCZ} \alpha_{\Xi}$
in parentheses. The statistical error in the MC sample is estimated by
varying the size of the MC sample used in the fit $(\pm 2\%)$.  The
error associated with the uncertainty in the $\Xi_c^0$ fragmentation
function ~\cite{XICFRAG} \footnote{Based on a measurement of the
  $\Xi_c^0$ fragmentation function at CLEO, we use a Peterson
  fragmentation function ~\cite{PETER} with $\epsilon_q = 0.15$.} is
estimated by varying this function $(+1\%,-14\%)$.  To determine the
effect of incomplete knowledge of the background shape and the effect of
statistical fluctuations in the sideband sample used to model the
background in the signal region, we vary the size of the lower and upper
sidebands used in the loglikelihood fit. We also repeat the fit using
wrong sign $\Xi^- \pi^-$ events in both the signal mass region and
sideband regions to model the background shape $(+16\%,-1\%)$. The error
associated with MC modeling of slow pions from $\Xi$ and $\LA$ decay is
obtained by varying the reconstruction efficiency according to our
understanding of the CLEO II detector $(+9\%,-6\%)$. Using a large
sample of $\CASCZHD$ MC generated with $\alpha_{\CASCZ} = -0.5$ and the
world average values of $\alpha_{\Xi}$ and $\alpha_{\LA}$ and including
a randomly generated background based on the shape of the data sideband,
we measure the effect of varying the size of the area element used to
determine $\Gamma_S(\cos \theta_{\Xi}, \cos \theta_{\LA};
\alpha_{\CASCZ}, \alpha_{\Xi}, \alpha_{\LA})$ and the background shape
$\Gamma_B(\cos \theta_{\Xi}, \cos \theta_{\LA})$ in the loglikelihood
fit $(+1\%,-2\%)$. The fit method is checked by integrating over
$\cos\theta_{\LA}$ ($\cos \theta_\Xi$) and performing
a one dimensional binned fit to $\cos \theta_\Xi$ ($\cos
\theta_{\LA}$). The results
are consistent with the maximum likelihood fit. Possible background 
from $\Sigma_c^0 (2450) \To \LCP
\pi^-_{\rm{slow}}$, $\LCP \To \LA \pi^+_{\rm {fast}}$ is determined to
be negligible by MC simulation.  This measurement is insensitive to
production polarization, ${\bf P_{\CASCZ}}$, and no
systematic error has been included from this source~\cite{ABC}.  Adding
all sources of systematic error in quadrature $(+19\%,-15\%)$, we find
$\alpha_{\CASCZ} \alpha_{\Xi} = \SLOPEAA$. A similar study for the
systematic uncertainty on the measurement of $\alpha_{\Xi} \alpha_{\LA}$
in $\CASCZHD$ yields $(-8\%,+4\%)$ and consequently $\alpha_{\Xi}
\alpha_{\LA} = \SLOPEBA$.

From the measurement of $\alpha_{\CASCZ} \alpha_{\Xi}$ and the PDG
evaluation: $\alpha_{\Xi} =-0.456 \pm 0.014$ ~\cite{PDG}, we obtain
$\AXCA$.  The physically allowed range of a decay asymmetry parameter is
$-1<\alpha<+1$. Our result prefers a negative value: $\alpha_{\CASCZ}$
is $<0.1$ at the 90\% CL.  The central value is in the middle of the
theoretically expected range but is not yet precise enough to choose
between models.

We note that $\alpha_{\Xi}$ has, so far, always been
determined as a product of decay asymmetries using Equation~\ref{lamasym}. 
In principle, given sufficient statistics, the 
loglikelihood method in conjunction with the two-fold joint angular
distribution of Equation~\ref{fullasym} allows the direct measurement
of all three asymmetry parameters: $\alpha_{\CASCZ}$, $\alpha_{\Xi}$,
and $\alpha_{\LA}$.

In conclusion, from a sample of $138 \pm 14$ $\CASCZHD$ decays 
we have measured $\alpha_{\CASCZ} \alpha_{\Xi} = \SLOPEAA$, from which
we obtain $\AXCA$.  To the best of our
knowledge this is the first measurement of a charm strange baryon decay
asymmetry parameter.


We gratefully acknowledge the effort of the CESR staff in providing us
with excellent luminosity and running conditions.  This work was
supported by the National Science Foundation, the U.S. Department of
Energy, Research Corporation, the Natural Sciences and Engineering
Research Council of Canada, the A.P. Sloan Foundation, the Swiss
National Science Foundation, and the Alexander von Humboldt Stiftung.

\end{document}